\documentclass[prb,aps,twocolumn,showpacs,bibnotes,superscriptaddress,epsf]{revtex4}

\usepackage{graphicx}
\usepackage{amsmath}
\usepackage{amssymb}
\usepackage{color}
\usepackage{dcolumn}
\usepackage{epsfig}
\usepackage{bm}
\usepackage{subfigure}
\usepackage[urlcolor=blue]{hyperref}
\hypersetup{backref, colorlinks=true, linkcolor=blue, citecolor=blue}

\begin{document}

\title{Observation of Meissner effect in potassium-doped \emph{p}-quinquephenyl}

\author{Ge Huang}
\affiliation{Center for High Pressure Science and Technology Advanced Research, Shanghai 201203, China}

\author{Ren-Shu Wang}
\affiliation{Center for High Pressure Science and Technology Advanced Research, Shanghai 201203, China}
\affiliation{School of Materials Science and Engineering, Hubei University, Wuhan 430062, China}

\author{Xiao-Jia Chen}
\email{xjchen@hpstar.ac.cn}
\affiliation{Center for High Pressure Science and Technology Advanced Research, Shanghai 201203, China}

\date{\today}

\begin{abstract}
The chain-like organic compounds with conjugated structure have the potential to become high temperature superconductors. We examine this idea by choosing \emph{p}-quinquephenyl with five phenyl rings connected in \emph{para} position. The \emph{dc} magnetic susceptibility measurements provide solid evidence for the presence of Meissner effect when the compound is doped by potassium. The real part of the \emph{ac} susceptibility shows exactly same transition temperature as that in \emph{dc} magnetization, and the imaginary part  of nearly zero value after transition implies the realization of zero-resistivity. All these features support the existence of superconductivity with a critical  temperature of 7.3 K in this material. The occurrence of bipolarons revealed by Raman spectra guarantees potassium metal intercalated into \emph{p}-quinquephenyl and suggests the important role of this elementary excitation played on superconductivity.
\end{abstract}
\pacs{74.70.-b, 74.20.Mn, 82.35.Lr, 78.30.Jw}

\maketitle

\section {Introduction}

Organic superconductors provide abundant basis to understand fundamental properties of correlated electrons.\cite{45} The large Coulomb correlation among conduction electrons, known as one of characterizations of organic superconductors, seems to be distinguished from the electron-phonon coupling in the conventional Bardeen-Cooper-Schrieffer (BCS) theory.\cite{2a,2b} Furthermore, the critical temperatures (\emph{T}$_c'$s) of these organic superconductors were predicted to be dramatically high because of high electron-electron interaction energy as discussed by V. L. Ginzburg and W. A. Little, respectively.\cite{a,b} So far, superconductivity in these materials emerges with expense of some competing orders, such as charge density wave, spin density wave, anti-ferromagnetism, and so on.\cite{c,d} The characteristic of these competing orders in the normal state of organic superconductors has been studied extensively.\cite{46,e,47,48} Organic metals were first observed to exhibit superconductivity in quasi-one-dimensional (TMTSF)$_2$PF$_6$ by the use of pressure to overcome spin density wave order.\cite{e} Antimagnetic order is also illustrated to compete with superconductivity in the $\kappa$-(BEDT-TTF)$_2$X family (X=Cu[N(CN)$_2$]Cl, Cu[N(CN)$_2$]Br, and Cu(NCS)$_2$)\cite{f} and the cubic alkali-metal doped fullerene Cs$_3$C$_{60}$.\cite{g}

In accordance to organic metals, conducting polymers, such as polyparaphenylene (PPP), possess low-dimensional nature and large Coulomb correlation. Polymer conductors also exhibit high electrical conductivities upon doping with donors or acceptors.\cite{3a} PPP is of chain-like structure with infinite benzene rings linked with C-C single bond. It shows typical nondegenerate ground-state property. Oligophenyls as models of PPP were found to undergo structure modification upon cooling by heat capacity measurement, X-ray diffraction, and Raman scattering measurement.\cite{j,k,l} Comparing with original unit cell, the lattice constant almost double in the \emph{b} and \emph{c} directions.\cite{m} The softening phonon mode\cite{l} in Raman spectra is exactly the same as the behaviors of the formation of charge-density-wave order in transition metal dichalcogenides, such as NbSe$_2$.\cite{49,50,51} The charge carriers in doped polymers are considered to be paramagnetic polarons (S=1/2) and diamagnetic bipolarons (S=0) with charge of -\emph{e} (or +\emph{e}) and -2\emph{e} (or +2\emph{e}). The electronic state and structural changes over several units are involved in the formation of polarons or bipolarons.\cite{h} When dopants are taken into account, bipolarons are proved to be more stable excitation with negative effective correlation energy \emph{U}$_{eff}$, the difference between the Coulomb repulsion and the lattice relaxation energy.\cite{i} The characteristic of bipolarons is analogous to the Cooper pair in the BCS theory of superconductivity. All these features make conducting polymers as the promising candidates for high-temperature superconductors. Recently, \emph{p}-terphenyl as a member of oligophenyls was pronounced to become superconductors upon doping potassium metal with \emph{T$_c$} at 7.2 K, 43 K and 123 K, respectively.\cite{n,o,p} This chain-like \emph{p}-terphenyl differs from the polycyclic aromatic hydrocarbons, such as phenanthrene,\cite{40,42} chrysene,\cite{42} picene,\cite{7} coronene,\cite{43} and 1,2;8,9-dibenzopentacene,\cite{44} due to the absence of zigzag and armchair edges. A superconducting phase with \emph{T$_c$} as high as 107 K was reported in K-doped \emph{p}-terphenyl flake as well.\cite{5b} Furthermore, the photoemission spectroscopy conducted in surface K-doped \emph{p}-terphenyl crystals shows a gap persisting up to at least 60 K and shares similar temperature dependence with obtained spectra of BSCCO superconductors.\cite{q} An analogous gap below 50 K was also found in K-doped \emph{p}-terphenyl films fabricated on Au (111), the stubborn feature of this gap at the applied magnetic field up to 11 T indicates the high upper critical field for such a kind of high-\emph{T$_c$} superconductor.\cite{4b}

Both the higher doping level and longer chain length are favor of huge increase of electrical conductivity for PPP.\cite{r} We expect the existence of superconductivity in \emph{p}-quinquephenyl consisting of five benzene rings linked in the para position. \emph{p}-Quinquephenyl, which can be seen as the most ideal, simple, rigid and symmetric molecule, has attracted increasing attention as liquid crystal.\cite{26,28} Meanwhile, \emph{p}-quinquephenyl can be exploited in organic field-effect transistors because of its electrical properties.\cite{27} We synthesize the K-doped \emph{p}-quinquephenyl and observe the Meissner effect in this material. Both \emph{dc} and \emph{ac} magnetic measurements support the existence of superconductivity in this material. Raman scattering measurements reveal the occurrence of bipolarons. The observation of superconductivity in K-doped \emph{p}-quinquephenyl adds a new member of oligophenyls superconductors after \emph{p}-terphenyl.

\section{Experimental Details}

Potassium metal (99\% purity) and \emph{p}-quinquephenyl (98\% purity) were both purchased from Alfa Aesar. The synthesis method was reported earlier.\cite{n} We got the black powder sample of K-doped \emph{p}-quinquephenyl after sealing and annealing in quartz tube under high vacuum. The quartz tube was then heated to 380 $^\circ$C in 70 minutes, and held at 340 $^\circ$C for 3 days. The resulting black powder samples, K$_x$\emph{p}-quinquephenyl were completely different from the white pristine \emph{p}-quinquephenyl. Potassium metal is easy to get oxidized when it is exposed to the air. All these experiments excluding annealing were done in glove box(O$_2$ less than 0.1 ppm, H$_2$O less than 0.1 ppm). The doped sample was sealed into nonmagnetic capsules for magnetization and Raman scattering measurements. Magnetization measurements of both the \emph{dc} and \emph{ac} susceptibility were carried out on doped samples as functions of temperature and field by using a SQUID magnetometer (Quantum Design MPMS3). The temperature ranges from 1.8 K to 20 K and the external magnetic field is applied up to 1000 Oe. Raman spectroscopy measurements were performed on K$_x$\emph{p}-quinquephenyl and pristine \emph{p}-quinquephenyl in backscattering geometry by using an in-house system with Charge Coupled Device and Spectrometer from Princeton Instrument. The laser excitation is in a wavelength of 660 nm and power less than 1 mW to avoid damage to samples.

\begin{figure}[tbp]
\includegraphics[width=\columnwidth]{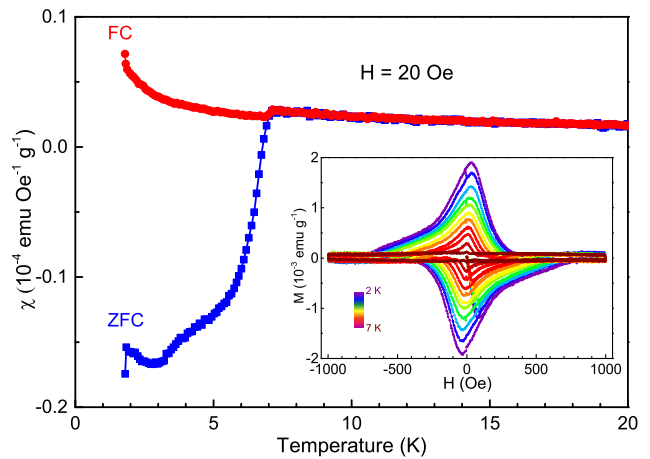}
\caption{Temperature dependence of the \emph{dc} magnetic susceptibility for K$_x$\emph{p}-quinquephenyl in a low magnetic field of 20 Oe, within the zero-field cooling and the field cooling cycles. Inset shows the magnetic hysteresis with applied field up to 1000 Oe at various temperatures below 7.3 K.}
\end{figure}

\section{Results and Discussion}

The Meissner effect and zero resistivity are two fundamental features of superconductivity. Actually the metal-doped carbon-based superconductors were mainly detected from magnetic measurement.\cite{7,40,42,43,44} However, with small diamagnetic volume and easily damaged crystal structure under pressure, the metallic state of K$_x$\emph{p}-quinquephenyl was hard to be realized even below \emph{T$_c$}. Difficulties in detecting weak superconductivity from resistivity measurements can be overcome by using magnetic susceptibility experiments.

Figure 1 shows the \emph{dc} magnetic susceptibility $\chi$ for the powder sample of K$_x$\emph{p}-quinquephenyl versus temperature measured in the zero-field cooling (ZFC) and field cooling (FC) cycles at 20 Oe. It should be pointed out that the $\chi$ in the ZFC run drops sharply below 7.3 K and then saturates. At around 5 K, the second step transition originates in the possibility for another superconducting phase. Meanwhile, the $\chi$ in the FC run decreases slowly and the saturation value is much smaller, suggesting existence of pining due to magnetic impurity. The temperature corresponding to the sharp drop is defined as \emph{T}$_c$. The shielding fraction extracted from the ZFC $\chi$ is only 0.074\%, if we assume a density of about 3 g$\cdot$cm$^{-3}$. Such a small shielding fraction is due to the presence of the impurities of this powder sample. The diamagnetic signal is at least an order of magnitude smaller than the signal obtained in bulk sample.\cite{7,6} Note that both ZFC and FC $\chi$ show an upturn trend at a lower temperature than \emph{T}$_c$, indicating the strong paramagnetic signal. The inset of Fig. 1 shows the M versus H plots obtained at different temperatures ranging from 2 K to 7 K with scanning field up to 1000 Oe after removing the paramagnetic backgrounds. The M(H) curves become totally nonlinear upon 100 Oe so that the lower critical field is close to 100 Oe. Nevertheless, the magnetization of K$_x$\emph{p}-quinquephenyl does not reach zero in applied fields up to 1000 Oe, implying the upper critical field is much higher than 1000 Oe. This diamond-like hysteresis loop is a typical character for the type-II superconductor.

\begin{figure}[tbp]
\includegraphics[width=\columnwidth]{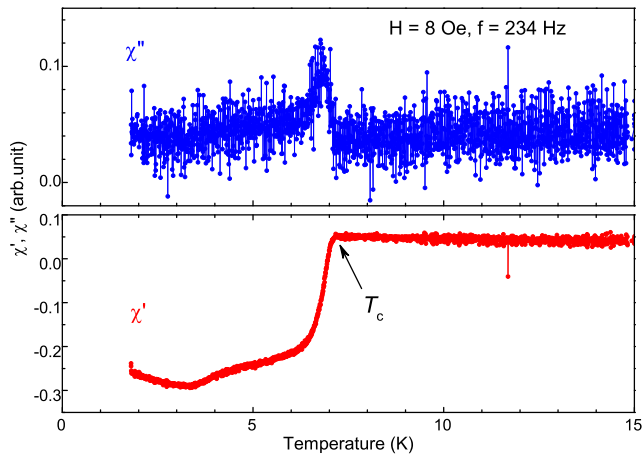}
\caption{Temperature dependence of the real (${\chi'}$) and imaginary (${\chi''}$) parts of the \emph{ac} magnetic susceptibility of K$_x$\emph{p}-quinquephenyl.}
\end{figure}

Furthermore, we have also measured \emph{ac} magnetic magnetization to confirm the existence of superconductivity in K$_x$\emph{p}-quinquephenyl. The complex \emph{ac} magnetic susceptibility denoted as: $\chi$ = $\chi$$'$ + i$\chi$$''$, has been successfully used to determine some parameters of superconductivity, such as the critical temperature and magnetic fields, critical current density, London and Campbell penetration depths, the pining potential, irreversibility line, and so on.\cite{4} Figure 2 shows the temperature dependence of the first harmonic \emph{ac} magnetic susceptibility for K$_x$\emph{p}-quinquephenyl in an external \emph{ac} magnetic field of 8 Oe and frequency at 234 Hz. The observation of double step-like transition in the real part of susceptibility $\chi$$'$ can be ascribed to the screening properties and the granular nature of our samples.\cite{5} The imaginary part of susceptibility $\chi$$''$ exhibits a positive peak, indicating the penetration of the \emph{ac} field into the sample. The \emph{T}$_c$ has been assigned to the separation point between $\chi$$'$ and $\chi$$''$. After this point, the almost zero signal of $\chi$$''$ upon cooling implies the realization of zero-resistivity in the superconductivity state.\cite{n,4,6}

Superconductivity in K$_x$\emph{p}-quinquephenyl has been detected by both the \emph{dc} and \emph{ac} magnetic measurements. The \emph{ac} magnetic measurements have been widely used in cuperate and iron-based superconductors. However, the \emph{ac} magnetic measurements are absent in most polycyclic aromatic hydrocarbon superconductors.\cite{62} The resistivity measurements have been applied in organic salts to identify the superconducting phase, but the zero-resistivity in superconducting state was rarely achieved.\cite{d,2,3}

\begin{figure}[tbp]
\includegraphics[width=\columnwidth]{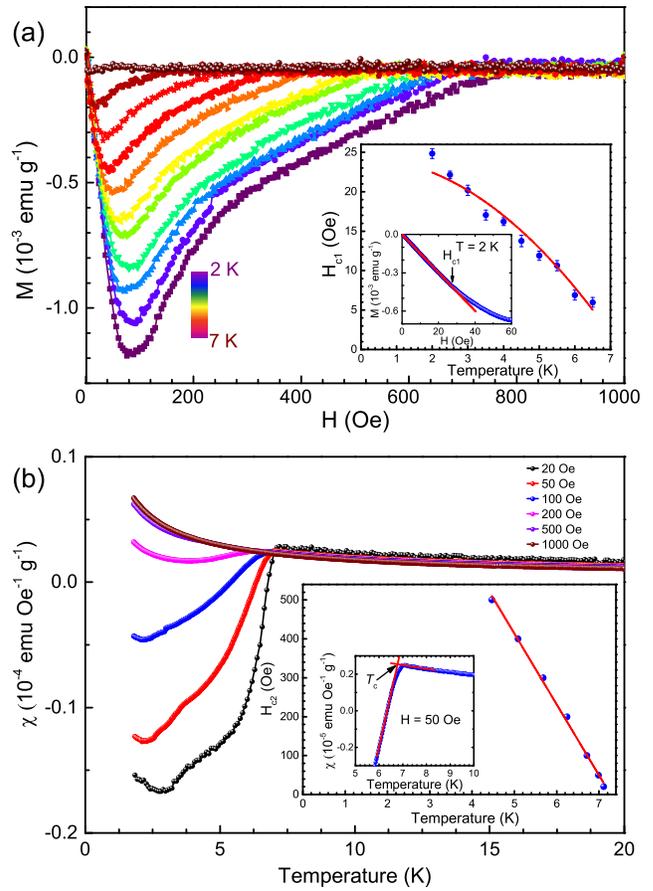}
\caption{(a) The initial part of M-H curves measured at low fields at various temperature. Inset: The temperature dependence of lower critical field \emph{H}$_{c1}$(T) and the determination of \emph{H}$_{c1}$ at temperature of 2 K given by the field dependent magnetization. (b) The temperature dependence of magnetic susceptibility for K$_x$\emph{p}-quinquephenyl obtained at different magnetic fields in the zero-field cooling cycles. Inset: The temperature dependence of the upper critical field \emph{H}$_{c2}$(T) and an example used to definite \emph{T}$_c$ by the magnetization versus temperature at the field of 50 Oe.}
\end{figure}

The magnetic field dependence of the magnetization for potassium-doped \emph{p}-quinquephenyl at various temperatures with low applied fields is summarised in Fig. 3(a). The linear behavior of magnetic-dependent magnetization signals the Meissner effect in this superconductor. The \emph{H}$_{c1}$ at selected temperatures is defined as the terminative point of transition from linear to nonlinear M(H) as shown in the smallest inset of Fig. 3(a). The method used to determine \emph{H}$_{c2}$ is similar to the previous studies.\cite{8} The regression coefficient R of a linear fit to the data points collected between 0 and H, is used as a function of H. The \emph{H}$_{c1}$ is determined based on the deviation from linear dependence of the R versus H. The lower critical fields \emph{H}$_{c1}$ at given temperatures are shown in the inset and the solid line represents the empirical law \emph{H}$_{c1}$(\emph{T})/\emph{H}$_{c1}$(0) = 1-(\emph{T}/\emph{T$_c$})$^2$. The lower critical field at zero-temperature \emph{H}$_{c1}$(0) is 24.2$\pm$0.6 Oe.

Figure 3(b) shows the temperature dependence of the magnetic susceptibility of our sample measured under applied fields in the ZFC run. The diamagnetic volume becomes smaller with increasing magnetic fields and \emph{T}$_c$ is thus decreased slowly towards lower temperature. It seems likely that the superconducting fraction is totally suppressed by applied field at 1000 Oe. However, due to the existence of strong paramagnetic impurities, the diamagnetic signal is covered by the strong background. The inset of Fig. 3(b) exhibits that the intercept of linear extrapolations from below and above superconducting transition is defined as \emph{T$_c$}. As an example, the result measured at field of 50 Oe is shown. The upper critical field \emph{H}$_{c2}$ can not be definitely determined from these magnetic measurements.\cite{7} The roughly estimation of \emph{H}$_{c2}$ as a function of temperature is shown in the inset of Fig. 3(b). The straight line is the data fitting from the Werthamer-Helfand-Hohenberg formula.\cite{9} We obtain the upper critical field at zero-temperature \emph{H}$_{c2}$(0)=1324.3$\pm$24.4 Oe.

The zero-temperature superconducting London penetration depth $\lambda$$_L$ and Ginzburg-Landau coherence length $\xi$$_{GL}$ can be estimated from \emph{H}$_{c1}$(0) and \emph{H}$_{c2}$(0) by using the equation \emph{H}$_{c2}$(0)=$\Phi$$_0$/2$\pi$$\xi$$_{GL}^2$ and \emph{H}$_{c1}$(0)= ($\Phi$$_0$=4$\pi$$\lambda$$_L^2$)$\ln$($\lambda$$_L$=$\xi$$_{GL}$),\cite{61} and the quantum flux $\Phi$$_0$ is $\phi_0$ = 2.0678$\cdot$10$^{-15}$ Wb. $\xi$$_{GL}$ of 498.6 ${\AA}$ and $\lambda$$_L$ of 520.8 ${\AA}$ are obtained for this \emph{p}-quinquephenyl superconductor from substituting \emph{H}$_{c1}$(0) and \emph{H}$_{c2}$(0). The obtained Ginzburg-Landau parameter $\kappa$ = $\lambda_L$/$\xi_{GL}$ = 1.04 also suggests that the K$_x$\emph{p}-quinquephenyl belongs to a type-\uppercase\expandafter{\romannumeral2} superconductor.

\begin{figure}[tbp]
\includegraphics[width=\columnwidth]{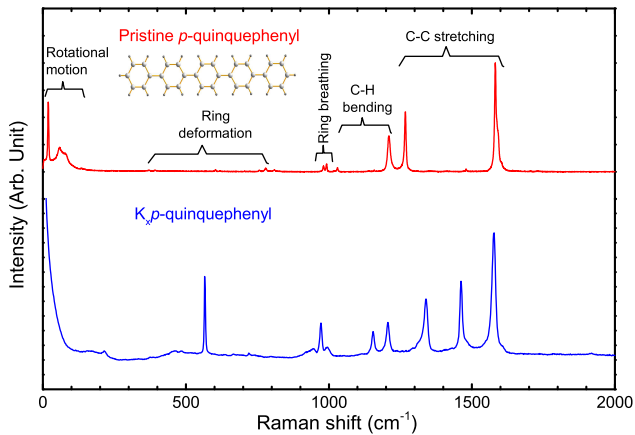}
\caption{Raman spectra of pristine \emph{p}-quinqquephenyl and K$_x$\emph{p}-quinquephenyl measured at room temperature.}
\end{figure}

Figure 4 shows Raman spectra of pure \emph{p}-quinquephenyl and potassium-doped \emph{p}-quinquephenyl. \emph{p}-Quinquephenyl has a benzenoid structure with five benzene rings aligned in para position, as shown in Fig. 4. All of these major peaks in pristine \emph{p}-quinquephenyl can be assigned and classified as five parts: rotational motion, ring deformation, ring breathing, C-H bending and C-C stretching.\cite{19,20}

The bipolaronic characterization in alkali-metal doped polyparaphenylenes synthesized by different method has been extensively studied in previous works.\cite{21,22} The almost separated two intra-ring C-C stretching modes at around 1582 cm$^{-1}$ in parent sample downshift and merge to bipolaronic bands localized at 1576 cm$^{-1}$. The observation of 1461 cm$^{-1}$ mode with no corresponding band in the pristine can be considered as the fingerprint for the formation of bipolarons. The band near 1268 cm$^{-1}$ assigned inter-ring C-C stretching upshifts to 1338 cm$^{-1}$ by intercalated potassium metal into \emph{p}-quinquephenyl, indicating that the chain length become more coplanar and the lattice evolves from a benzoid structure to a quinoid structure.\cite{21} The 1154 cm$^{-1}$ and 1206 cm$^{-1}$ modes presented in Raman spectra of doped samples are due to C-H external and internal stretching, respectively. These two peak are both evolved from 1211 cm$^{-1}$ mode in the pristine. The triple bands localized at 970 cm$^{-1}$ are from the formation of bipolarons. Raman characterization of potassium-doped \emph{p}-quinquephenyl proves the existence of bipolarons, which possibly accounts for the observed superconductivity in K-doped \emph{p}-terphenyl.\cite{n,o,p}

\section{Conclusion}
In summary, a superconducting phase in K$_x$\emph{p}-quinquephenyl is identified by both the \emph{dc} and \emph{ac} magnetic measurements. The magnetic magnetization at various temperatures and fields reveals the Meissner effect in the superconducting state. The Ginzburg-Landau parameter $\kappa$=1.04 suggests that the K-doped \emph{p}-quinquephenyl is a type-\uppercase\expandafter{\romannumeral2} superconductor. We thus have discovered a new member of oligophenyls superconductors. Now that the bipolarons were proposed to account for the electrons pairing, it seems to be applied to superconductivity in K-doped \emph{p}-quinquephenyl as well.

We thank Guo-Hua Zhong, Hai-Qing Lin, Yun Gao, and Zhong-Bing Huang for valuable discussions.

\end{document}